\documentclass[10pt,letterpaper]{article}
\usepackage{opex3}

\begin{document}

\title{Bose-Einstein condensate in an optical lattice with tunable spacing:\\
transport and static properties}

\author{Leonardo Fallani, Chiara Fort, Jessica E. Lye and Massimo Inguscio}

\address{LENS European Laboratory for Nonlinear Spectroscopy,\\
Dipartimento di Fisica - Universit\`a di Firenze, and INFM,\\
via Nello Carrara 1, I-50019 Sesto Fiorentino (FI), Italy}

\email{fallani@lens.unifi.it}

\begin{abstract*}
In this Letter we report the investigation of transport and static
properties of a Bose-Einstein condensate in a large-spaced optical
lattice. The lattice spacing can be easily tuned starting from few
micrometers by adjusting the relative angle of two partially
reflective mirrors. We have performed \emph{in-situ} imaging of
the atoms trapped in the potential wells of a 20 $\mu$m spaced
lattice. For a lattice spacing of 10 $\mu$m we have studied the
transport properties of the system and the interference pattern
after expansion, evidencing quite different results with respect
to the physics of BECs in ordinary near-infrared standing wave
lattices, owing to the different length and energy scales.
\end{abstract*}


\section{Introduction}

In the last few years Bose-Einstein condensates (BECs) in optical
lattices have been the subject of extremely intense and rewarding
research, both theoretical and experimental. Periodic potentials
produced with near-infrared standing waves have been used to
investigate the transport and superfluid properties of
Bose-condensed samples \cite{anderson,science}, as well as to
study effects correlated with the physics of strongly correlated
many-body systems \cite{greiner1}. Ultracold atoms in optical
lattices are also good candidates for the implementation of
quantum computational schemes \cite{qcbrennen,qcjaksch}. In this
respect, quantum logic operations have already been performed via
cold controlled collisions on entangled sets of atoms trapped in a
three-dimensional optical lattice \cite{mandel}. In a recent
experiment a one-dimensional lattice partially filled with atoms
has been used to create a quantum register with the capability of
single atom manipulation and detection \cite{meschede}. In this
context, a larger spacing optical lattice could simplify the
fundamental goal of single-site addressability, that is hard to
achieve in a traditional near-infrared standing-wave lattice. One
possible approach consists in the realization of standing waves
with CO$_2$ lasers emitting at 10 $\mu$m wavelength \cite{co2}.
Even if CO$_2$ lattices provide the advantage of a complete
suppression of heating mechanisms (due to the huge detuning from
the atomic resonance), they present the drawback of a difficult
experimental manipulation. A different approach involves the
realization of arrays of dipole traps obtained by shining a far
detuned laser beam onto a microfabricated array of microlenses
\cite{ertmer}. Optical lattices with a spacing of few $\mu$m can
also be obtained from the interference of two near-infrared laser
beams intersecting at a small angle
\cite{arimondo,phillips,dalibard}. In the frame of this idea, in
this Letter we present a simple system for the creation and the
detection of optical lattices with spacing starting from $\approx
8$ $\mu$m, almost 20 times larger than the spacing of a
traditional standing wave lattice, still using near-infrared light
produced by solid state laser sources. The appeal of this system
resides in the possibility to easily detect the lattice intensity
profile and to accomplish a fine tuning of the lattice spacing
simply by adjusting the relative angle of two partially reflective
mirrors. We have investigated both static and dynamic properties
of an $^{87}$Rb BEC loaded in such a potential, evidencing quite
different results with respect to the physics of ordinary
lattices, caused by the different length and energy scales.

\section{Experimental setup}

\begin{figure}[t!]
\begin{center}
\includegraphics[width=0.7\columnwidth]{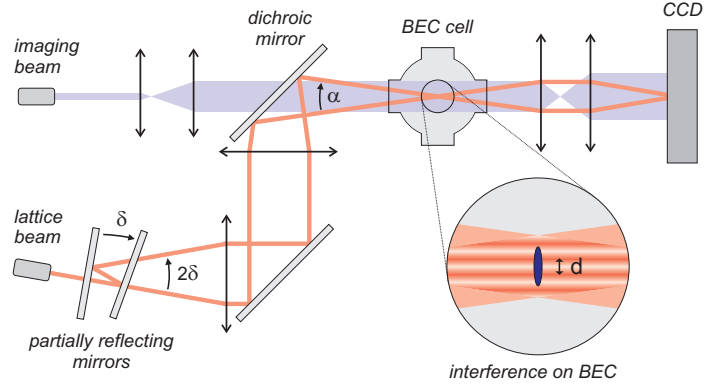}
\end{center}
\caption[Large-spacing optical lattice setup]{Optical setup for
the production and detection of a large spacing optical lattice.
The two beams creating the interference pattern are obtained from
the multiple reflections of a laser beam by a pair of partially
reflective mirrors placed with a relative angle one in front of
the other. The lattice spacing $d$ can be tuned by changing the
angle $\delta$ between the two partially reflecting mirrors. The
vertical axis is orthogonal to the page.} \label{fig01}
\end{figure}

The experimental setup for the production and the detection of the
large spacing optical lattice is schematically shown in Fig.
\ref{fig01}. A collimated laser beam coming from a Ti:Sa laser is
shone onto a pair of partially reflective mirrors placed with a
small relative angle $\delta$ one in front of the other at a
distance of $\approx 1$ mm. As shown in Fig. \ref{fig01}, the
multiple reflection of the laser beam from these mirrors produces,
at the second order of reflection, two separate beams (of
different intensities) with a relative angle $2 \delta$. These two
beams, following different optical paths, are then guided by a
lens system to recombine onto the condensate, where they interfere
producing a periodic pattern with alternating intensity maxima and
minima. The period $d$ of this lattice, that is oriented along the
difference of the wavevectors, depends on the angle $\alpha$
between the beams according to
\begin{equation}
d = \frac{\lambda}{2 \sin(\alpha/2)}.
\end{equation}
It is easy to show that in the case of counterpropagating beams
($\alpha=\pi$) the above expression reduces to the well known
spacing $\lambda/2$ for a standing wave lattice. In the other
limit, when the two beams are almost copropagating ($\alpha \simeq
0$), $d$ may become very large. In our setup, varying the angle
$\delta$ between the two mirrors, it is possible to easily adjust
the lattice spacing to the desired value. For the working
wavelength $\lambda = 820$ nm the lower limit is $d \approx 8$
$\mu$m, corresponding to the maximum angle $\alpha = 25^\circ$
that is possible to reach in our setup taking into account the
finite size of the vacuum cell windows. Since the two beams
producing the lattice do not have the same intensity, we expect
the resulting interference pattern to show a reduced contrast with
respect to the case in which the two beams have the same
intensity. It can be shown that the intensity of the lattice
$I_L$, i.e. the intensity difference between constructive
interference and destructive interference, is given by
\begin{equation}
I_L =I_{max}-I_{min}=4 t^2 \left( 1 - t \right) I_0,
\end{equation}
where $t$ is the transmissivity of the partially reflecting
mirrors and $I_0$ is the intensity of the beam incident on them.
It is easy to show that this expression has a maximum for $t=2/3$.
For this reason in the experiment we have used mirrors with $t
\simeq 0.7$, corresponding to the maximum lattice intensity $I_L
\simeq 0.59 I_0$ achievable with this technique.

An appealing feature of the system shown in Fig. \ref{fig01} is
that the lattice beams are aligned quasi-parallel to the radial
horizontal axis of the condensate, following the same path of the
imaging beam used in our setup. This has been made possible by the
use of a dichroic mirror, reflecting in the range $\lambda > 800$
nm and transmitting in the range $\lambda < 800$ nm. This feature
allows us to use the same imaging setup to detect both the BEC and
the spatial profile of the lattice light intensity. This means
that we can image in consecutive photos both the condensate and
the exact potential that the condensate experiences. Indeed, since
the CCD plane is conjugate to the vertical plane passing through
the trap axis, the intensity profile recorded by the CCD is
exactly the same (except for a magnification factor) as the one
imaged onto the condensate. Furthermore, by calibrating the CCD
responsivity with a reference beam of known intensity, it is
possible to convert the digitized signal of each pixel into an
intensity value and thus calculate the height of the potential
$V_0$. In the following, the lattice height and the other energy
scales will be conveniently expressed in frequency units (using
the implicit assumption of a division by the Planck constant $h$).

\begin{figure}[t!]
\begin{center}
\includegraphics[width=0.8\columnwidth]{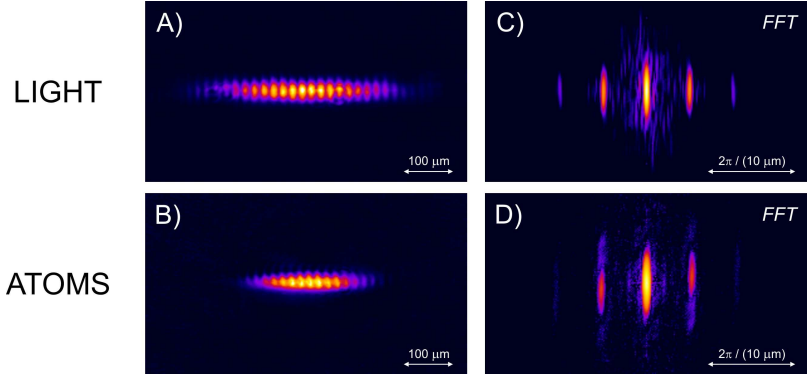}
\end{center}
\caption[Optical lattice with 20 $\mu$m spacing]{An optical
lattice with $d=20$ $\mu$m spacing. In the top row we show the
intensity distribution of the light recorded by the CCD (a) and
the corresponding Fourier transform (c). In the bottom row we show
the density distribution (b) of the atoms trapped in the lattice
sites imaged \emph{in situ} a few $\mu$s after switching off the
lattice, together with its Fourier transform (d), showing well
resolved peaks in the same position as the ones in (c).}
\label{fig02}
\end{figure}

To illustrate the advantages of this imaging setup, in Fig.
\ref{fig02}(a) we show the intensity profile of an optical lattice
with spacing $d=20$ $\mu$m, while in Fig. \ref{fig02}(b) we show
an absorption image of the atoms trapped in the potential wells of
the same lattice. The experimental sequence used to trap the atoms
in the optical lattice and to image \emph{in situ} the atomic
distribution is the following. First we produce a cigar-shaped BEC
of $^{87}$Rb in a Ioffe-Pritchard magnetostatic trap, with trap
frequencies $\omega_z / 2 \pi = 8.74(1)$ Hz along the symmetry
axis $z$ (oriented horizontally) and $\omega_r / 2 \pi = 85(1)$ Hz
along the orthogonal directions. The typical diameter of the
condensates is 150 $\mu$m axially and 15 $\mu$m radially. We note
that the lattice beam profile has been tailored with cylindrical
lenses in order to match the condensate elongated shape (see Fig.
\ref{fig02}(a)). After producing the BEC, still maintaining the
magnetic confinement, we ramp in 100 ms the height of the lattice
from zero to the final value by using an acousto-optic modulator
(AOM). After the end of the ramp we wait 50 ms, then we abruptly
switch off the optical lattice with the same AOM and, after a few
tens of $\mu$s, we flash the imaging beam for the detection phase.
The latter time interval is necessary not to perturb the imaging
with lattice light coming onto the CCD, but is small enough not to
let the atoms expand from the lattice sites once the optical
confinement is released. The combination of the CCD electronic
shutter and an interferential bandpass filter placed in front of
the camera (centered around $\lambda = 780$ nm) allows a complete
extinction of the lattice light at the time of acquisition. In
Fig. \ref{fig02}(c) and \ref{fig02}(d) we show, respectively, the
power spectrum of the two-dimensional Fourier transform of the
distributions shown in Figs. \ref{fig02}(a) and \ref{fig02}(b). As
one can see, both the distributions are characterized by sharp
peaks in momentum space: from the position of these peaks it is
possible to precisely measure the spatial period of the observed
structures.

\begin{figure}[t!]
\begin{center}
\includegraphics[width=0.9\columnwidth]{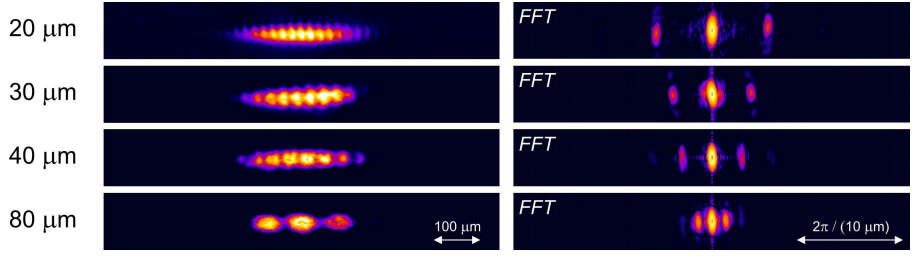}
\end{center}
\caption{\emph{In situ} images of the atoms trapped in the lattice
sites (left) and Fourier transform of the density distribution
(right) for different lattice spacings from 20 to 80 $\mu$m. The
different spacings have been obtained by changing the angle
$\delta$ between the mirrors shown in Fig. 1.} \label{fig03}
\end{figure}

In Fig. \ref{fig03} we show \emph{in situ} images of the atomic
sample for different lattice spacings ranging from 20 $\mu$m to 80
$\mu$m. The different spacings are obtained, as shown before, by
simply changing the angle $\delta$ between the mirrors shown in
Fig. \ref{fig01}.

\section{Static properties}

With the usual near-infrared standing-wave lattices employed so
far in many experiments it is not possible to optically resolve
\emph{in situ} the modulation of the atomic density distribution,
since the distance between lattice sites is typically less than 1
$\mu$m. As a matter of fact, in most of the cases the diagnostic
of the system is carried out by looking at the atomic gas after
expansion. This kind of analysis provides useful information on
the quantum nature of the system. When the height of the optical
lattice is larger than the chemical potential (tight binding
regime), the system forms an array of condensates localized in the
wells of the periodic potential. It has been experimentally
observed that such an array of coherent atomic states, once
released from the trap, produces after expansion a periodic
interference pattern \cite{greiner1,greiner2d,pedri}. The high
contrast interference observed in these experiments has been
related to the long range coherence that exists in the superfluid
regime when the tunnelling rate between neighboring sites is
sufficiently large. Since the tunnelling rate strongly depends on
the lattice spacing $d$ (being proportional to $e^{-2d}/\sqrt{d}$
in the tight binding regime \cite{zwerger}), it is worth studying
the expansion of the BEC from a large-spaced lattice, in which
tunnelling is expected to be exponentially suppressed.

Let us consider a linear array of condensates trapped in an
optical lattice with spacing $d$. After releasing the atoms from
the trapping potential one expects to observe a periodic
interference pattern with spacing
\begin{equation}
d'=\frac{h t_{exp}}{md},
\end{equation}
where $m$ is the atomic mass, $t_{exp}$ is the expansion time and
$d$ is the spacing of the optical lattice \cite{ketterleinterf}.
We have studied the expansion from a lattice with $d=10$ $\mu$m,
producing after $t_{exp}=28$ ms an interference pattern with
fringe spacing $d'= 12.8$ $\mu$m, easily detectable with our
imaging setup. We note that for this lattice spacing the recoil
energy $E_R=h^2 /8md^2$, the natural energy scale for measuring
the lattice height, is only 6 Hz, almost 600 times smaller than
the recoil energy for a regular standing-wave lattice with $d=0.4$
$\mu$m spacing.

\begin{figure}[t!]
\begin{center}
\includegraphics[width=0.8\columnwidth]{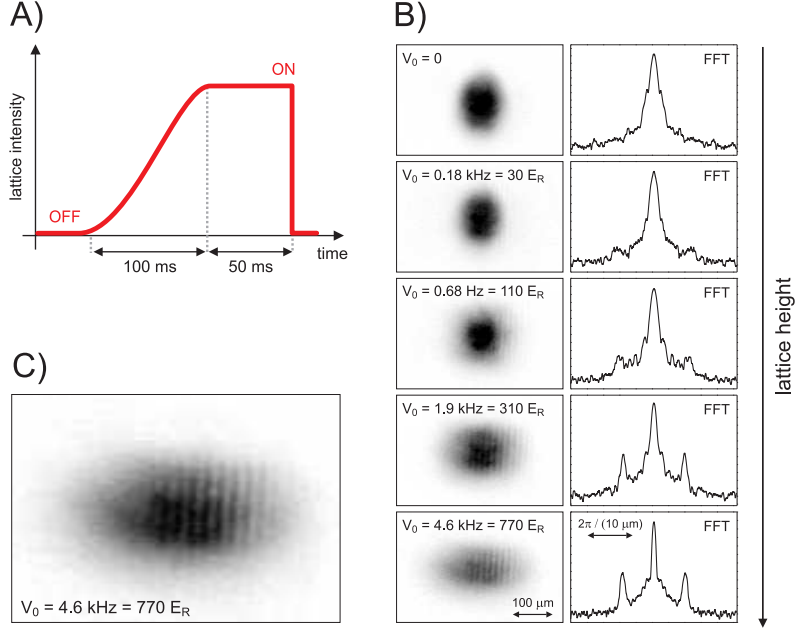}
\end{center}
\caption[Expansion from a 10 $\mu$m lattice]{Expansion from an
optical lattice with 10 $\mu$m spacing. a) The intensity of the
lattice is increased adiabatically from zero to the final value in
100 ms, then after 50 ms the lattice is abruptly switched off
together with the harmonic trapping potential. b) Absorption
images after 28 ms of expansion (left) and corresponding Fourier
transform (right) for different lattice heights $V_0$. c) Expanded
density profile for $V_0=4.6$ kHz: we observe the clear presence
of interference fringes.} \label{fig04}
\end{figure}

In the experiment, after producing the BEC, we ramp in 100 ms the
intensity of the lattice beam from zero to different final values,
then we wait 50 ms and suddenly switch off both the magnetic trap
and the optical lattice, as indicated in Fig. \ref{fig04}(a). In
Fig. \ref{fig04}(b) we show absorption images of the atomic
density distribution after 28 ms of free expansion (left) together
with their Fourier transform evaluated along the lattice axis
(right) for different values of the lattice height ranging from
$V_0=0$ to $V_0=4.6$ kHz. Increasing the lattice height above
$\sim 500$ Hz we note that interference fringes with the expected
spacing $d' \simeq 13$ $\mu$m start to form, as indicated by the
emergence of peaks in the Fourier transform, and their visibility
increases with increasing lattice height. At the same time, the
axial width of the overall distribution gets larger as a
consequence of the increased axial confinement in the lattice
wells, that produces a faster expansion along the lattice
direction. In Fig. \ref{fig04}(c) we show the interference pattern
observed for the maximum lattice height $V_0 = 4.6$ kHz, in which
the system is deeply in the tight binding regime and we expect the
BEC to be split into an array of condensates located in the
lattice wells (the BEC chemical potential in the harmonic trap is
$\mu \simeq 1$ kHz). However, differently from \cite{pedri}, for
this large spacing the tunnelling between neighboring sites is
totally suppressed and the states in the different wells do not
communicate one with each other. As a consequence, each state will
evolve in time independently, according to its energy, that is
different from site to site due to the inhomogeneity of the
sample. In this situation we are observing interference fringes
from an array of phase uncorrelated matter-wave sources, as
recently reported in \cite{dalibard} with a similar system.

\begin{figure}[t!]
\begin{center}
\includegraphics[width=0.75\columnwidth]{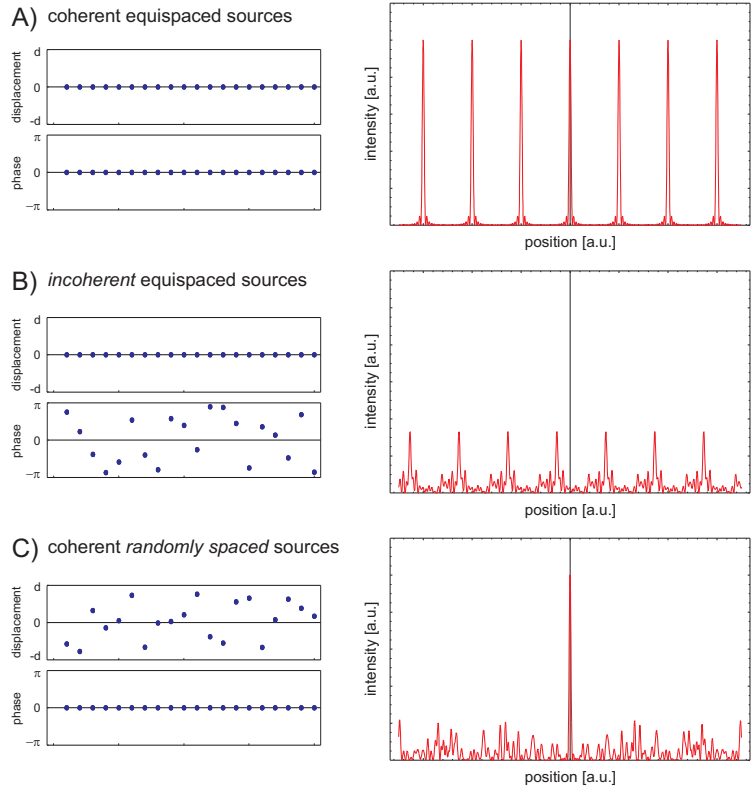}
\end{center}
\caption[Interference from point-like sources]{Far-field intensity
obtained from the interference of a linear chain of 20 point-like
emitters. On the left of each row the diagrams show the particular
set of positions and phases of the sources used to calculate, with
Eq. (\ref{eq:interf0}), the interferograms shown on the right. The
three sets refer to: a) uniform phase and uniform spacing; b)
random phase and uniform spacing; c) uniform phase and random
spacing.} \label{fig05}
\end{figure}

To get more insight into the problem of interference from
uncorrelated sources, let us write down a very simple model.
Following an obvious analogy with optics, we consider a linear
array of point-like radiation sources disposed along $\hat{z}$
with uniform spacing $d$, all emitting isotropically in space with
the same amplitude. This problem is the extension of the Young's
double slit experiment to the case of $N$ emitters. We indicate
the position of the sources along $\hat{z}$ and their phases with
the variables $\{z_n,\phi_n\}$. Let us suppose that we measure the
field distribution on a screen parallel to $\hat{z}$, placed at a
distance $D$ from the emitters. The field amplitude, as a function
of the position $z'$ on the screen, is proportional to a sum of
phase-factors describing the wave propagation in space:
\begin{equation}
A(z') \propto \sum_n e^{i\left( \phi_n + k d_n\right)},
\end{equation}
where $k$ is the modulus of the wavevector and
$d_n=\sqrt{D^2+(z_n-z')^2}$ is the distance of the $n$-th source
from the detection point. If one assumes that $D\gg (z_n-z')$,
i.e. that we are observing the interference in the far-field, we
can make the following approximation
\begin{equation}
d_n =\sqrt{D^2+(z_n-z')^2} \simeq D \left[ 1 + \frac{1}{2} \left(
\frac{z_n-z'}{D} \right)^2 \right].
\end{equation}
Using this assumption, the total intensity on the screen will be
given by
\begin{equation}
\left| A(z') \right|^2 \propto \left| \sum_n e^{i\left( \phi_n +
\frac{k}{2D} (z_n-z')^2\right)} \right|^2. \label{eq:interf0}
\end{equation}
This quantity is evaluated numerically in Fig. \ref{fig05} for
$N=20$ sources identified by three different sets of variables
$\{z_n,\phi_n\}$, in which we release one at a time the hypothesis
of identical phases and uniform spacing. The diagrams on the left
show the displacement of each source $\delta z_n=z_n-nd$ from the
regular lattice position and its phase $\phi_n$. The graphs on the
right show the field intensity $\left| A(z') \right|^2$ calculated
with Eq. (\ref{eq:interf0}).

In Fig. \ref{fig05}(a) we consider the ideal case of equispaced
sources all emitting in phase. In this situation we observe a
high-contrast interference pattern characterized by well resolved
peaks at a distance inversely proportional to the distance $d$
between the sources. In optics, this is the intensity distribution
produced by a diffraction grating illuminated by coherent light.
In matter-wave optics, a similar interferogram is observed in the
superfluid regime after the expansion of BECs released from
optical lattices produced with near-infrared standing waves
\cite{pedri}. Working out a slightly more realistic model, in
which the finite size of the emitters is taken into account, the
interferogram should be convolved with the diffraction figure from
a single source, resulting in a decreased visibility of the higher
order peaks.

In Fig. \ref{fig05}(b) we consider the case of equispaced sources
with random phases. The interferogram on the right corresponds to
the randomly generated set of phases shown on the left. In this
case, even if no coherence is present across the array, we can
still observe a periodic structure in the interferogram. This is
indeed what we observe in Fig. \ref{fig04}(c) and what has been
studied, both theoretically and experimentally, in
\cite{dalibard}. We note that this periodic interference is
produced in a single shot. Actually, averaging the intensity
distribution over many different realizations, the contrast of the
interferogram is expected to rapidly vanish, since the
uncorrelated phases produce a pattern that differs from shot to
shot both in the relative position of the peaks and in their
visibility. The persistence of a periodic interference pattern
even in the case of phase-uncorrelated sources is an effect
essentially related to the limited number of emitters that
interfere. Indeed, in the case of only two sources, even if there
is no phase relation between them, one expects to observe a
perfect interference pattern with $100 \%$ contrast
\cite{ketterleinterf}. Increasing the number of sources, the
visibility of the peaks decreases and more complex structures
start to grow. Even if the overall distribution still shows
periodicity, the harmonic content of the interferogram grows at
the expense of a reduced visibility and, in the limit of a very
large number of sources, the interference pattern could be
confused with noise. However, a correlation measurement should
still be able to detect a periodic structure, that is closely
related to the ordered distribution of the emitters. A detailed
analysis of such a problem is carried out in \cite{dalibard}.

In Fig. \ref{fig05}(c) we consider the case of coherent sources
with a random displacement from their regular position $z_n=nd$.
The interferogram on the right corresponds to the randomly
generated set of displacements shown on the left. While in the
case of random phases and uniform spacing a periodic interference
pattern is still visible in a single shot, in the case of random
spacing no characteristic structure is visible at all, even if all
the sources emit coherently. This model reproduces the density
distribution of a BEC released from a deep optical speckle
potential, where a broad gaussian density profile, without any
internal structure, has been experimentally observed
\cite{random}.

\section{Transport properties}

\begin{figure}
\begin{center}
\includegraphics[width=0.75\columnwidth]{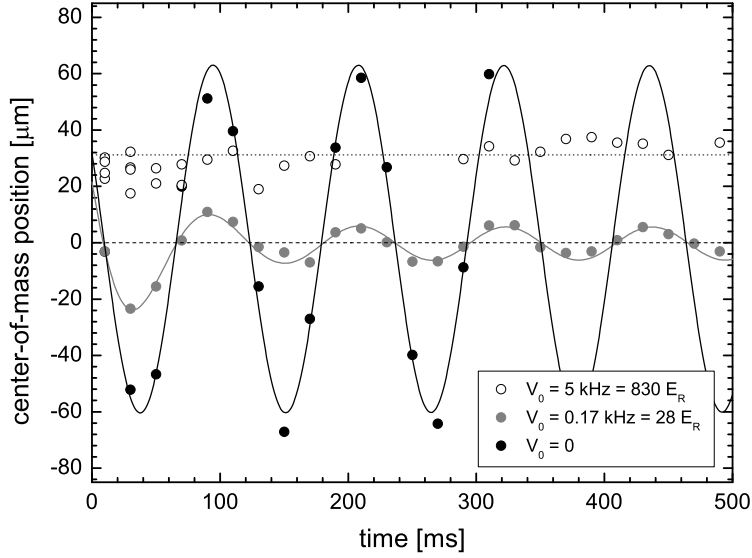}
\end{center}
\caption[Dipole oscillation in a 10 $\mu$m lattice]{Dipole
oscillations of a harmonically trapped BEC in the presence of an
optical lattice with 10 $\mu$m spacing. The dotted line is the
initial position of the atoms at $t=0$. The dashed line is the
center of the magnetic trap after the excitation of the dipole
mode. The black points refer to the oscillation without lattice,
the gray points show a damped oscillation in a lattice with height
$V_0=170$ Hz, while the empty circles show the localization in a
lattice with height $V_0=5$ kHz.} \label{fig06}
\end{figure}

In order to check the actual localization of the system in the
regime of large lattice heights in which we observe interference
fringes, we have studied dipole oscillations of the harmonically
trapped BEC in the presence of the lattice. In order to induce
dipole oscillations we suddenly shift the center of the magnetic
trap over a distance $\Delta z$ along the trap axis, as described
in \cite{science} with more detail. In Fig. \ref{fig06} we show
the center of mass position of the expanded cloud as a function of
time for a trap displacement $\Delta z=32$ $\mu$m and different
lattice heights. The black points refer to the regular undamped
oscillation of the condensate in the pure harmonic potential at
the trap frequency $\nu_z = (8.74 \pm 0.03)$ Hz.

\begin{figure}
\begin{center}
\includegraphics[width=0.75\columnwidth]{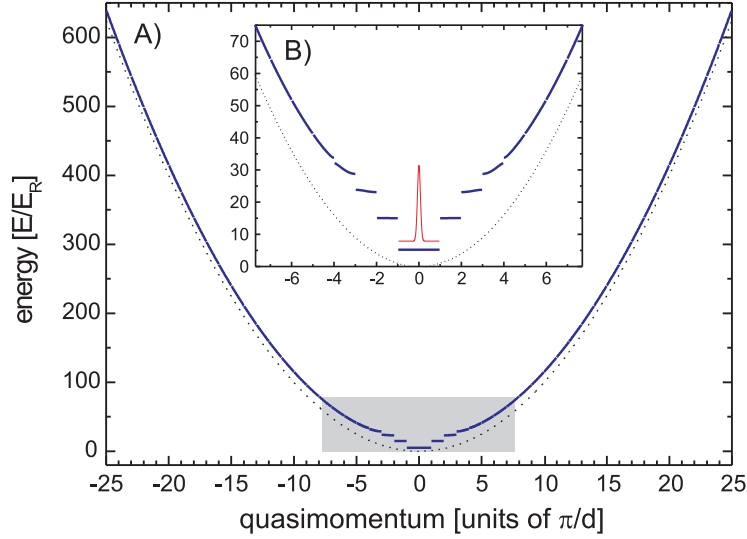}
\end{center}
\caption{a) Band structure for a quantum particle in a periodic
potential with spacing $d=10$ $\mu$m for a lattice height
$V_0=28E_R$ (solid line) and $V_0=0$ (dotted line). The width of
the graph ($50\pi/d$) corresponds to the extension of the first
Brillouin zone for a regular lattice with $0.4$ $\mu$m spacing. b)
Zoom of the gray box. The width of this graph is the actual range
of momenta spanned by the atoms during the oscillation in the
harmonic potential for the trap displacement $\Delta z = 32$
$\mu$m used in the experiment. The thin line represents the
extension of the atomic wavepacket in momentum space.}
\label{fig07}
\end{figure}

The gray points correspond to the center-of-mass motion in a
shallow lattice with $V_0 = 170$ Hz $=28 E_R$. In this case we
observe a damped oscillation of the center of mass at the same
frequency of the harmonic trap. This result is qualitatively
different from what reported in \cite{science}, where the authors
evidenced a frequency shift of the dipole oscillations in a
standing wave lattice with $d\simeq0.4$ $\mu$m. This shift was
explained in terms of a modified effective mass $m^*$ at the
bottom of the energy band, which changes the frequency from
$\nu_z$ to $\sqrt{m/m^*}\nu_z$. Here the main difference is that,
for a lattice spacing $d=10$ $\mu$m and a trap displacement
$\Delta z = 32$ $\mu$m, the motion of the atomic cloud in momentum
space is no longer confined in the center of the first Brillouin
zone, where the band is approximately parabolic. Let us assume
that the Bloch picture is still valid in this regime, in which the
condensate occupies only $\sim 15$ sites. In Fig. \ref{fig07}(a)
we show the band structure for a quantum particle in a periodic
potential with spacing $d=10$ $\mu$m. The solid line refers to a
lattice height $V_0=28E_R$, while the dotted line shows the free
particle parabolic spectrum. The width of the graph ($50\pi/d$)
corresponds to the extension of the first Brillouin zone for a
regular lattice with $0.4$ $\mu$m spacing. Since now the lattice
spacing is 25 times bigger, the corresponding width of the
Brillouin zones $\pi/d$ is 25 times smaller. Indeed, for the
oscillation amplitudes we are considering, the atoms are spanning
many Brillouin zones during their motion, making Landau-Zener
tunnelling from one band to the next one at each passage from the
zone edges. As a consequence, the actual energy spectrum probed by
the atoms during the oscillation is basically the free particle
one, with a curvature set by the real mass $m$, except for a small
modification at the bottom, where small energy gaps form between
the very first bands. The width of the gray region represents the
actual range of momenta spanned by a particle oscillating in the
pure harmonic trap for an initial displacement $\Delta z=32$
$\mu$m. In Fig. \ref{fig07}(b) we show a zoom of the gray region,
in which the thin continuous line represents the extension of the
atomic wavepacket in momentum space (its rms width being
$0.07\pi/d$). In order to observe effects of reduced effective
mass, like the ones studied in \cite{science}, one should use a
much smaller amplitude oscillation ($\sim 1$ $\mu$m), that could
be barely detectable with our imaging setup, in order to stay
within one Brillouin zone. The observed damping of the
center-of-mass oscillation could be attributed to the finite
probability of interband Landau-Zener tunnelling during the motion
of the wavepacket through the periodic potential. The splitting of
the wavepacket into multiple bands then results in a dephasing of
the different states leading to a damping of the center-of-mass
motion. In addition, the finite transverse size of the lattice
beam could play a role, causing the height of the optical barriers
to be slightly different from site to site, thus breaking the
exact translational symmetry of the lattice potential. In order to
better understand which of these two mechanisms is more relevant
for the observed dynamics further studies have to be carried out.

Coming back to Fig. \ref{fig06}, increasing the height of the
optical lattice to $V_0 = 5$ kHz $=830 E_R$, we observe that the
center-of-mass motion is blocked and the atomic cloud stays at a
side of the displaced harmonic trap. Indeed, in this regime the
height of the periodic potential becomes larger than the chemical
potential of the condensate and the BEC is split into an array of
condensates located at the different sites. Differently from the
standing wave lattice \cite{science}, in which the tunnelling
ensures a collective motion even in the tight binding regime, in
this case the tunnelling between adjacent sites is heavily
suppressed by the large distance between wells, and the atomic
states in each site are expected not to communicate one with each
other. We note however that, even for this lattice height, we
detect clear interference fringes as shown in Fig. \ref{fig04}(c)
for all the evolution times considered in Fig. \ref{fig06}. The
hypothesis of independence of the condensates localized at the
lattice sites formulated before is thus confirmed by the
observation that, in the presence of an external potential
gradient, the center of mass does not move in time.

We would like to note that this localization effect cannot be
attributed to the generation of a Mott insulator state
\cite{greiner1}, as also suggested in \cite{dalibard}. Indeed, in
the Mott insulator phase the system presents nontrivial
localization properties on a timescale much longer than the
tunnelling times. In our case the tunnelling is so heavily
suppressed that the observation of the system is necessarily
limited to a timescale on which the long-range coherence
properties of the system could not be detected. Actually, what we
are observing in the experiment is a localization effect produced
only by the strongly increased tunnelling times, and not by an
actual competition between interaction and tunnelling energies
\cite{greiner1}.

\section{Conclusions}

In this Letter we have presented a simple system for the
production and detection of large-spacing optical lattices with
adjustable spatial period. We have investigated transport and
static properties of a Bose-Einstein condensate loaded in a
periodic potential with 10 $\mu$m spacing. In particular, we have
evidenced the presence of interference fringes after expansion
even in the insulating regime in which inter-well tunnelling is
heavily suppressed and the center-of-mass dynamics is inhibited.
The appeal of this system is that, by increasing the lattice
spacing to 20 $\mu$m or more, it becomes possible to optically
resolve \emph{in situ} the single lattice sites. This possibility
could be important for the implementation of quantum computing
schemes, where addressability is a fundamental requirement. As an
extension of this work, the following step could be made in the
direction of manipulating the single sites either optically or
with the application of radiofrequency/microwave transitions
coupling different internal levels.

\section*{Acknowledgements}

This work has been supported by the EU Contracts No.
HPRN-CT-2000-00125, INFM PRA ``Photon Matter'' and MIUR FIRB 2001.
We thank all the people of the Quantum Gases group in Florence for
stimulating discussions, in particular Michele Modugno for
theoretical support and Vera Guarrera for careful reading of the
manuscript.

\end{document}